\documentclass[aps,prl,reprint,showpacs,groupedaddress]{revtex4-1}
\usepackage{amstext}
\usepackage{graphicx}
\usepackage{dcolumn}
\usepackage{bm}
\usepackage{hhline}

\begin{document}  
  
  \renewcommand{\andname}{\ignorespaces}

\title{Highly-sensitive superconducting circuits at $\sim$700$\,$kHz with tunable quality factors for image-current detection of single trapped antiprotons}

\author{H. Nagahama$^{1,2}$, G. Schneider$^{1,3}$, A. Mooser$^{1}$, C. Smorra$^{1,4}$, S. Sellner$^{1}$,  J. Harrington$^{5}$, T. Higuchi$^{1,2}$, M. Borchert$^{6}$, T. Tanaka$^{1,2}$, M. Besirli$^{1}$, K. Blaum$^{5}$, Y. Matsuda$^{2}$, C. Ospelkaus$^{6,7}$, W. Quint$^{8}$, J. Walz$^{3,9}$, Y. Yamazaki$^{10}$, and S. Ulmer$^{1}$}
\affiliation{$^1$ RIKEN, Ulmer Initiative Research Unit,  2-1 Hirosawa, Wako, Saitama 351-0198, Japan}
\affiliation{$^2$ Graduate School of Arts and Sciences, University of Tokyo, Tokyo 153-8902, Japan}
\affiliation{$^3$ Institut f\"ur Physik, Johannes Gutenberg-Universit\"at 55099 Mainz, Germany}
\affiliation{$^4$ CERN, 1211 Geneva, Switzerland}
\affiliation{$^5$ Max-Planck-Institut f\"ur Kernphysik, Saupfercheckweg 1, 69117 Heidelberg, Germany}
\affiliation{$^6$ Institut f\"ur Quantenoptik, Leibniz Universit\"at Hannover, Welfengarten 1, 30167 Hannover, Germany}
\affiliation{$^7$ Physikalisch-Technische Bundesanstalt, Bundesallee 100, 38116 Braunschweig, Germany}
\affiliation{$^8$ GSI - Helmholtzzentrum f\"ur Schwerionenforschung GmbH, 64291 Darmstadt, Germany}
\affiliation{$^9$ Helmholtz-Institut Mainz, 55099 Mainz, Germany}
\affiliation{$^{10}$ RIKEN, Atomic Physics Research Unit, 2-1 Hirosawa, Wako, Saitama 351-0198, Japan}


\begin{abstract}
We developed highly-sensitive image-current detection systems based on superconducting toroidal coils and ultra-low noise amplifiers for non-destructive measurements of the axial frequencies (550$\sim$800$\,$kHz) of single antiprotons stored in a cryogenic multi-Penning-trap system. The unloaded superconducting tuned circuits show quality factors of up to $500\,000$, which corresponds to a factor of 10 improvement compared to our previously used solenoidal designs. Connected to ultra-low noise amplifiers and the trap system, signal-to-noise-ratios of 30$\,$dB at quality factors of $>20\,000$ are achieved. In addition, we have developed a superconducting switch which allows continuous tuning of the detector's quality factor, and to sensitively tune the particle-detector interaction. This allowed us to improve frequency resolution at constant averaging time, which is crucial for single antiproton spin-transition spectroscopy experiments, as well as improved measurements of the proton-to-antiproton charge-to-mass ratio.
\end{abstract}

\maketitle

\section{I. Introduction}
Experiments with single particles in Penning traps allow measurement of particle properties with highest precision and give access to stringent tests of fundamental symmetries \cite{ref1}. Measurements in this field rely, to a large extent, on the non-destructive detection of the trapped particle's eigenfrequencies. For instance, the most precise measurements of atomic masses are based on Penning trap experiments \cite{E.Myers}. These measurements have impact on neutrino physics \cite{He-3,ref2}, as well as on stringent tests of relativistic energy-mass equivalence \cite{ref3}. Electron $g$-factor measurements provide the most stringent tests of bound-state quantum electrodynamics \cite{ref4}. From a $g$-factor measurement of the electron bound to a highly-charged carbon ion, the to-date the most precise value of the mass of the electron in atomic mass units has been extracted \cite{ref5}. Other Penning trap based measurements provide the most stringent tests of the fundamental charge, parity, time invariance. Electron/positron magnetic anomaly frequencies were compared with fractional precisions at the parts per billion (p.p.b$.$) level \cite{ref6}, and the fundamental properties of protons and antiprotons, such as charge-to-mass ratios \cite{ref7} and $g$-factors \cite{ref8} were compared with 90$\,$parts per trillion (p.p.t$.$) and 4.4$\,$parts per million (p.p.m$.$) precision, respectively. In our experiments we operate multi Penning-trap systems dedicated to comparing the fundamental properties of protons and antiprotons with unprecedented precision \cite{ref9}. Recently, we reported on the most precise comparison of the proton-to-antiproton charge-to-mass ratio with a fractional precision of 69$\,$p.p.t$.$ \cite{charge-to-mass ratio Nature}, and the most precise measurement of the magnetic moment of a single trapped proton with a fractional precision of 3.3$\,$p.p.b$.$ \cite{ref10}. \\
Our measurements rely on non-destructive measurements of the characteristic oscillation frequencies \cite{image current} of the trapped particles: the modified cyclotron frequency $\nu_{+}$, the magnetron frequency $\nu_{-}$, and the axial frequency $\nu_{z}$. An invariance theorem \cite{invariance} $\nu_{c}=\sqrt{\nu_{+}^2+\nu_{-}^2+\nu_{z}^2}$ relates these eigenfrequencies to the free cyclotron frequency $\nu_{c}=(qB_{0})/(2\pi m)$, and consequently to the fundamental properties of the particle; charge $q$ and mass $m$. $B_0$ is the magnetic field at the center of the trap.\\
The oscillatory motions of the particles induce tiny image currents $I_{\textrm{p}}$ on the trap electrodes. These currents are detected by means of highly-sensitive superconducting tuned circuits which act at their resonance frequencies as effective resistors $R_{\textrm{p}}$. The detectors are connected directly to the trap electrodes. The resulting voltage signals are then amplified and the time-transients are analyzed by using fast Fourier transform (FFT) spectrum analyzers. \\
\\
Here, we report on the development of highly-sensitive image-current detection systems based on superconducting toroidal coils and ultra-low noise cryogenic amplifiers. This allows for the non-destructive measurement of the axial frequency $\nu_{z}$ of the antiproton. A superconducting switch allows us to sensitively tune particle-detector interaction. This novel feature is crucial for fidelity tuning in single spin-transition spectroscopy experiments \cite{UlmerPRL,MooserPRL} and will be used in our planned high-precision antiproton magnetic moment measurements.

\section{II. Detection principle}
\begin{figure}[htbp]
	\begin{center}
		\includegraphics[width=8.5cm]{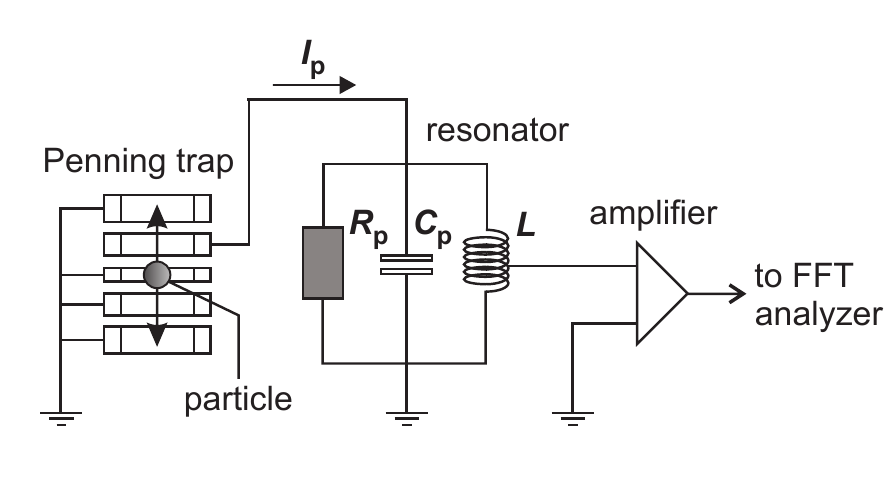}
		\caption{Schematic of detecting the axial frequency $\nu_{z}$ of a single trapped particle. The detector consists of a superconducting tuned circuit with resonance frequency matched to $\nu_{z}$ and a cryogenic ultra-low noise amplifier. The particle's eigenfrequency is obtained from the Fourier spectrum of the image current using an FFT analyzer. For details, see text.}
    \end{center}
\end{figure}
\begin{figure}[htbp]
	\begin{center}
		\includegraphics[width=8.5cm]{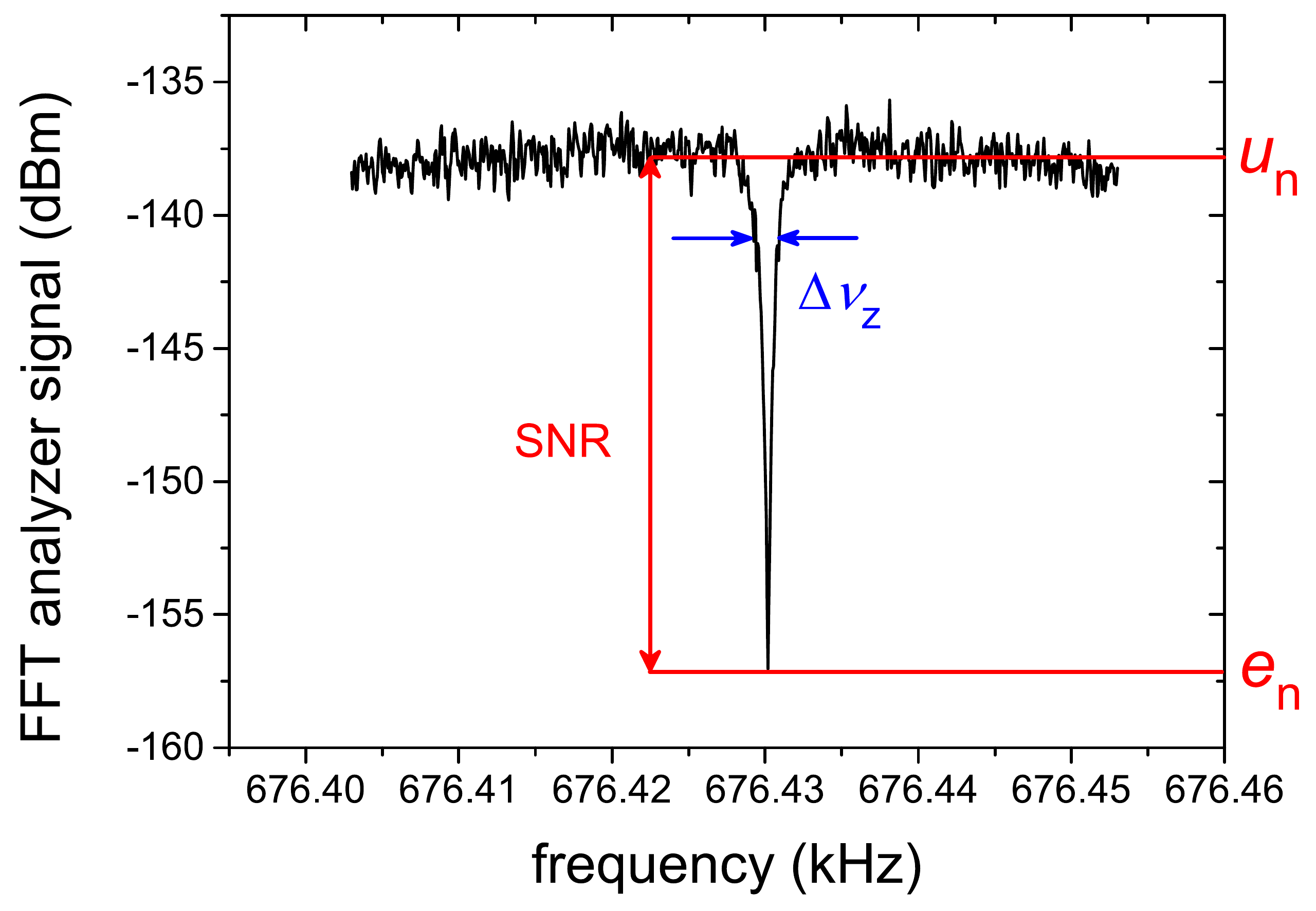}
		\caption{Example of a dip spectrum of a single antiproton stored in the precision trap. SNR is defined as the ratio of Johnson-Nyquist noise of the detector $u_{\textrm{n}}$ to the equivalent input noise of the amplifier $e_{\textrm{n}}$. $\Delta \nu_{z}$ is the 3$\,$dB width of the dip.}
	\end{center}
\end{figure}
FIG.$\,$1 shows a schematic for the detection of the axial frequency $\nu_{z}$ of a particle in a Penning trap. The detector consists of a superconducting tuned circuit with high quality factor ($Q$-value) and a cryogenic ultra-low noise amplifier.
The trap and resonator form a parallel $RLC$ tuned circuit, $L$ being the inductance of the superconducting coil and $C_{\textrm{p}}$ the sum of the parasitic capacitances of the trap and the coil itself. At resonance the reactances compensate each other and appear as an effective parallel resistance, $R_{\textrm{p}}=2\pi \nu_{0}LQ$.\\
The axial frequency of the trapped particle $\nu_{z}$ is tuned to the resonance frequency $\nu_{0}$ of the tuned circuit by adjusting the voltages applied to the Penning trap electrodes. A voltage drop $V_{\textrm{p}}=R_{\textrm{p}}I_{\textrm{p}}$ is detected, where
\begin{equation}
 I_{\textrm{p}}=\frac{2\pi q}{D}\times \nu_{z}z
\end{equation}
is the image current induced on the trap electrode, $z$ is the axial coordinate and $D$ is a trap-specific length, for the geometry of our traps, $D=11\,$mm. The signal $V_{\textrm{p}}$ is subsequently amplified by a cryogenic ultra-low-noise amplifier, and the time transient is recorded and processed by a fast Fourier transform (FFT) spectrum analyzer. Typical image currents are on the order of fA, therefore it is crucial to operate the detectors at high $R_{\textrm{p}}$. \\
When tuned to the resonance, the trapped particle is cooled resistively. Once cooled to thermal equilibrium, the particle acts as a series $LC$ circuit \cite{image current} and shorts the resonator's impedance. This produces a sharp dip at the eigenfrequency of the antiproton in the thermal noise spectrum of the detector, as shown in FIG.$\,$2. The signal-to-noise ratio (SNR) is the ratio of the Johnson-Nyquist noise of the detector $u_{\textrm{n}}$ to the equivalent input noise of the amplifier $e_{\textrm{n}}$, given by
\begin{equation}
	\textrm{SNR} = \frac{u_{\textrm{n}}}{e_{\textrm{n}}}\propto \frac{\sqrt{4k_{\textrm{B}}TR_{\textrm{eff}}}\cdot{\kappa}}{e_{\textrm{n}}},
\end{equation}
where $k_{\textrm{B}}$ is the Boltzmann-constant, $R_{\textrm{eff}}$ the effective parallel resistance of the whole system, $T$ the temperature of the detector, and $\kappa$ the coupling factor between tuned circuit and cryogenic amplifier. Note that high SNRs can either be achieved by optimizing $R_{\text{eff}}$ or by increasing the temperature of the system $T$, e.g. by coupling external noise to the circuit. However, the second approach will also increase particle temperature and therefore systematic shifts in frequency measurements \cite{charge-to-mass ratio Nature,ref10}, which is to be avoided.
By fitting a line-shape modeled dip to the measured spectrum, the axial frequency $\nu_{z}$ is obtained. The scatter $\sigma(\nu_{z,k}-\nu_{z,k-1})$ of a sequence of axial frequency measurements $\nu_{z,k}$ is given as
\begin{eqnarray}
	\sigma(\nu_{z,k}-\nu_{z,k-1})\propto\sqrt{\frac{1}{4\pi}\frac{\Delta \nu_z}{t_{\textrm{avg}}}\frac{1}{\sqrt{\textrm{SNR}}}},
\end{eqnarray}
where $t_{\textrm{avg}}$ is the averaging time and
\begin{eqnarray}
	\Delta\nu_z=\frac{1}{2\pi}\frac{R_{\textrm{eff}}}{m}\frac{q^2}{D^2}
\end{eqnarray}
the width of the particle dip. To enable fast particle detection with high frequency resolution, the SNR should be as large as possible. In addition, by tuning $R_{\textrm{eff}}$ and consequently $\Delta \nu_{z}$, the frequency resolution at constant averaging time can be improved.

\section{III. Resonator}
\begin{figure}[htbp]
	\begin{center}
		\includegraphics[width=8.5cm]{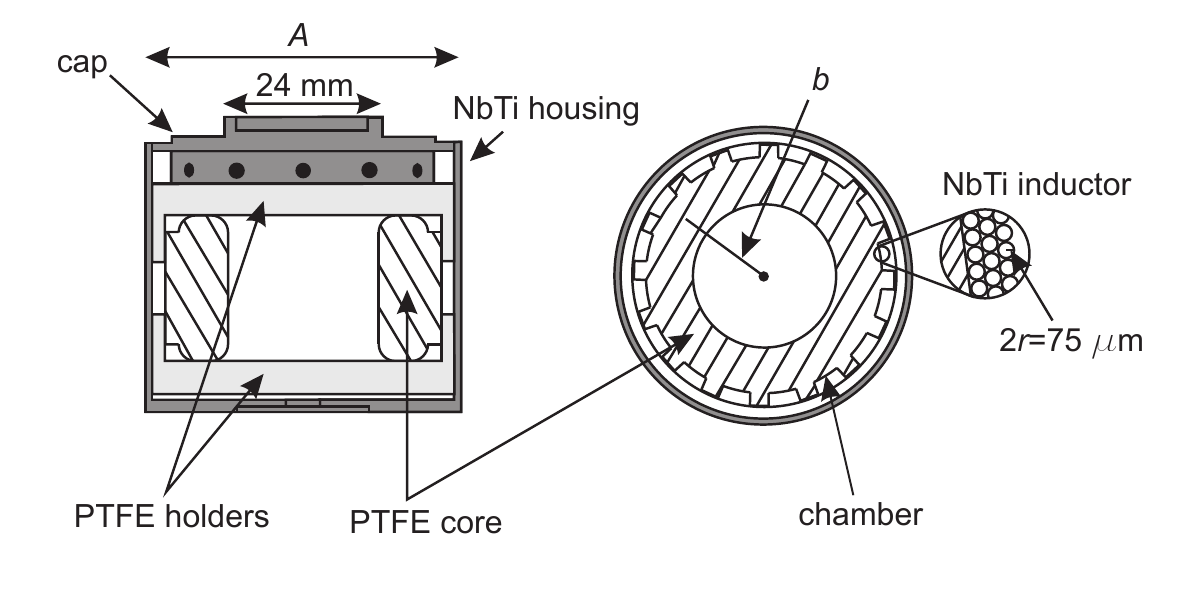}
		\caption{Sketches of the axial resonator. The toroidal core is made out of PTFE, and PTFE insulated NbTi superconducting wire is used for the windings. The toroid is mounted inside the NbTi housing and kept stable with PTFE holders. It is enclosed by inserting a cap from outside. Geometrical lengths differ for the two resonator designs.}
	\end{center}
\end{figure}
Each resonator consists of a toroidal coil in a cylindrical metal housing with an outer diameter $A$, as shown in FIG$.\,$3. The toroidal geometry confines the magnetic flux within the inductor, and consequently reduces losses caused by stray fields. The coils, as well as the housings, are made out of type-II superconducting NbTi. This material has a high critical magnetic field strength \cite{Bc2} of $B_{C2}=14.5\,$T, which enables placement of the resonators close to the trap system mounted in the high magnetic field $B_{0}=1.9\,$T of a superconducting magnet.\\
In total four resonators were developed, two with an outer diameter of $A$=48$\,$mm and two with $A$=41$\,$mm, the dimensions being defined by geometrical constraints of the experiment. The geometry of the toroid was optimized in a way, such that the inner cross-sectional area is maximized at the shortest length of superconducting wire, when considering the boundary conditions defined by the housings. To keep the parasitic capacitance of the coil small, we use three-layer chamber windings. The individual chambers are machined onto the toroid. Losses induced by dielectric polarization effects are kept small by making the cores of the toroids out of polytetrafluoroethylene (PTFE). At cryogenic temperatures and in our frequency range of interest, this material has a dielectric loss tangent of $\delta<0.0001$, which is supported by the experimental results described below.\\
For the superconducting wire, PTFE insulated NbTi wire with a diameter of 75$\,$\textrm{$\mu$}m is used. The windings are fixed to the core by PTFE thread-seal tape to ensure a good thermal contact between the wire and the PTFE core. This is crucial to allow appropriate thermalization. At each end of the coil, a 5$\,$cm long copper wire of 0.5$\,$mm diameter is connected. To keep resistive losses in the Nb/Ti-Cu joint small, we carefully degrease the respective metal contacts, place the ends in silver plated copper-ferrules add Pb/Sn solder and heat the joint slowly to 290$\, ^{\circ}$C. This toroid assembly is placed into PTFE holders, which are then mounted into the resonator housing. Finally, one end of the coil is directly soldered to ground (cold end), the other one is kept open, but efficiently thermalized by means of a custom-made sapphire capacitor. \\
For experimental reasons, we designed for different axial frequencies in different traps, and as a consequence different inductances $L$. In test experiments with copper wire we found that
\begin{equation}
	L=\frac{\mu_{0}S}{2\pi b}N^{2}
\end{equation}
reproduces the measured inductances within an error of 10$\,\%$. Here $\mu_{0}$ is the permeability constant of the vacuum, $S$ the cross-sectional area of a toroid, $b$ the radius of the central axis of a toroid, and $N$ the number of turns.\\
For characterization the coils are cooled to 4$\,$K and the $S_{21}$-transmission is measured using a vector network analyzer. The measurement setup is similar to the one described in our previous article \cite{Stefan_RSI}. From the transmission the resonance frequency $\nu_0$ and the 3$\,$dB width $\Delta\nu_r$ are determined and the quality factor is calculated as $Q=\nu_0/\Delta\nu_r$. Summarized results of these measurements are shown in TABLE$\,$I.\\
For all unloaded resonators we achieve quality factors $Q>190\,000$, corresponding to effective series resistances $R_{\rm{s}}<0.06\,\Omega$. These resistances may arise from residual losses in the dielectrics present in the resonator housing, as well as small residual resistances in the critical NbTi-to-Cu joints. However, the quality factors achieved here exceed the requirements of our trap-experiment by a factor of 10 and further optimization of the measured $R_{\textrm{s}}$ was not required. The obtained $Q$-values correspond to a factor of 10 improvement compared to our previously used solenoidal detection systems \cite{Stefan_RSI}.
\begin{table}[htb]
	\caption{Summary of the characterization measurements of the resonators. The left column indicates the name of the dedicated Penning trap in the BASE apparatus: the analysis trap (AT), cooling trap (CT), precision trap (PT), and reservoir trap (RT) \cite{ref9}. Abbreviations: $L$ - Inductance / $C_{\textrm{p}}$ - Parasitic capacitance / $\nu_{0}$ - Resonance frequency / $N$ - Number of turns / $R_{\textrm{p}}$ - Parallel resistance.}
  \begin{tabular}{|c||c|c|c|c|c|c|} \hline
     & $L$ & $C_{\textrm{p}}$ & $Q$-value & $\nu_{0}$ & $N$ & $R_{\textrm{p}}$ \\ \hhline{|=#=|=|=|=|=|=|}
    AT & 2.73$\,$mH & 9.5$\,$pF & 500$\,$000 & 896$\,$kHz & 1100 & 7.7$\,$G\textrm{$\Omega$} \\ \hline
    CT & 2.14$\,$mH & 11$\,$pF & 250$\,$000 & 948$\,$kHz & 940 & 3.2$\,$G\textrm{$\Omega$} \\ \hline
    PT & 1.75$\,$mH & 11$\,$pF & 194$\,$000 & 1.09$\,$MHz & 800 & 2.3$\,$G\textrm{$\Omega$} \\ \hline
    RT & 1.71$\,$mH & 11$\,$pF & 196$\,$000 & 1.07$\,$MHz & 800 & 2.3$\,$G\textrm{$\Omega$} \\ \hline
  \end{tabular}
\end{table}

\section{IV. Amplifier}
We use custom-made cryogenic amplifiers based on Dual-gate GaAs MES-FET to amplify the particle signals \cite{Stefan cyclotron}. They consist of a common-source circuit at the input and a source-follower circuit at the output. We use NE25139 (NEC) or 3SK164 (SONY) transistors for the input, and CF739 (Siemens / Tricomp) for the output stage. The parts are assembled on a high-quality PTFE based laminated printed circuit board material, which has a cryogenic loss-tangent \cite{PCB} of order tan$\, \delta \sim 10^{-4}$. This is important to prevent the reduction of the resonator $Q$-value by dielectric losses.\\
For the FETs used at the input stage, high effective input resistance $R_{\textrm{in}}$ and low equivalent input noise are crucial. At 4$\,$K, the equivalent input noise $e_{\textrm{n}}$ of the transistors at the resonance frequencies of the detectors is 680(30)$\,$pV$\cdot$Hz$^{-1/2}$. This was measured by using the noise marker function of a Rohde $\&$ Schwarz FSVR-spectrum analyzer. The equivalent input resistance $R_{\textrm{in}}$ is obtained by coupling the amplifier to a resonator with known quality factor, measuring loaded $Q$-value and using Eq.$\,$6 to determine $R_{\textrm{in}}=$8.5(5)$\,$M$\Omega$. A more detailed discussion of the amplifiers which are used in our experiment is described in reference \cite{Stefan cyclotron}.

\section{V. Coupling}
When coupling the amplifier with input resistance $R_{\textrm{in}}$ to the resonator, with its free equivalent parallel resistance $R_{\textrm{p}}$, the effective resistance $R_{\textrm{eff}}$ of the detection system becomes
\begin{eqnarray}
	R_{\textrm{eff}}=\frac{R_{\textrm{p}}R_{\textrm{in}}/\kappa^{2}}{R_{\textrm{p}}+R_{\textrm{in}}/\kappa^{2}}\,,
\end{eqnarray}
where
\begin{eqnarray}
	\kappa = \frac{C_{\textrm{a}}}{C_{\textrm{a}}+C_{\textrm{in}}} \times \frac{N_{2}}{N_{1}+N_{2}}
\end{eqnarray}
is the coupling factor. The term ${N_{2}}/({N_{1}+N_{2}})$ is adjusted by tapping the coil at a certain winding ratio, $N_2+N_1=N$ being the total number of windings of the coil. Additional capacitive decoupling is achieved by careful selection of the coupling capacitor $C_{\textrm{a}}$, and is used to fine-tune $\kappa$. \\
Depending on the requirements of the individual traps, we chose $\kappa$ by defining the target width of the axial frequency dip $\Delta\nu_z$, which is typically in a range between $1\,$Hz and $5\,$Hz. For traps which are used for precision frequency measurements, we chose the coupling factor $\kappa$ such that small dip widths of order 1$\,$Hz are obtained, for the other traps we chose values between 3$\,$Hz and 5$\,$Hz. By further decoupling of the amplifier, even higher $Q$ and $R_{\textrm{eff}}$ values can be achieved, however this is not meaningful in the context of our experiments.  

\section{VI. Implementation}
\begin{figure}[htbp]
	\begin{center}
		\includegraphics[width=8.5cm]{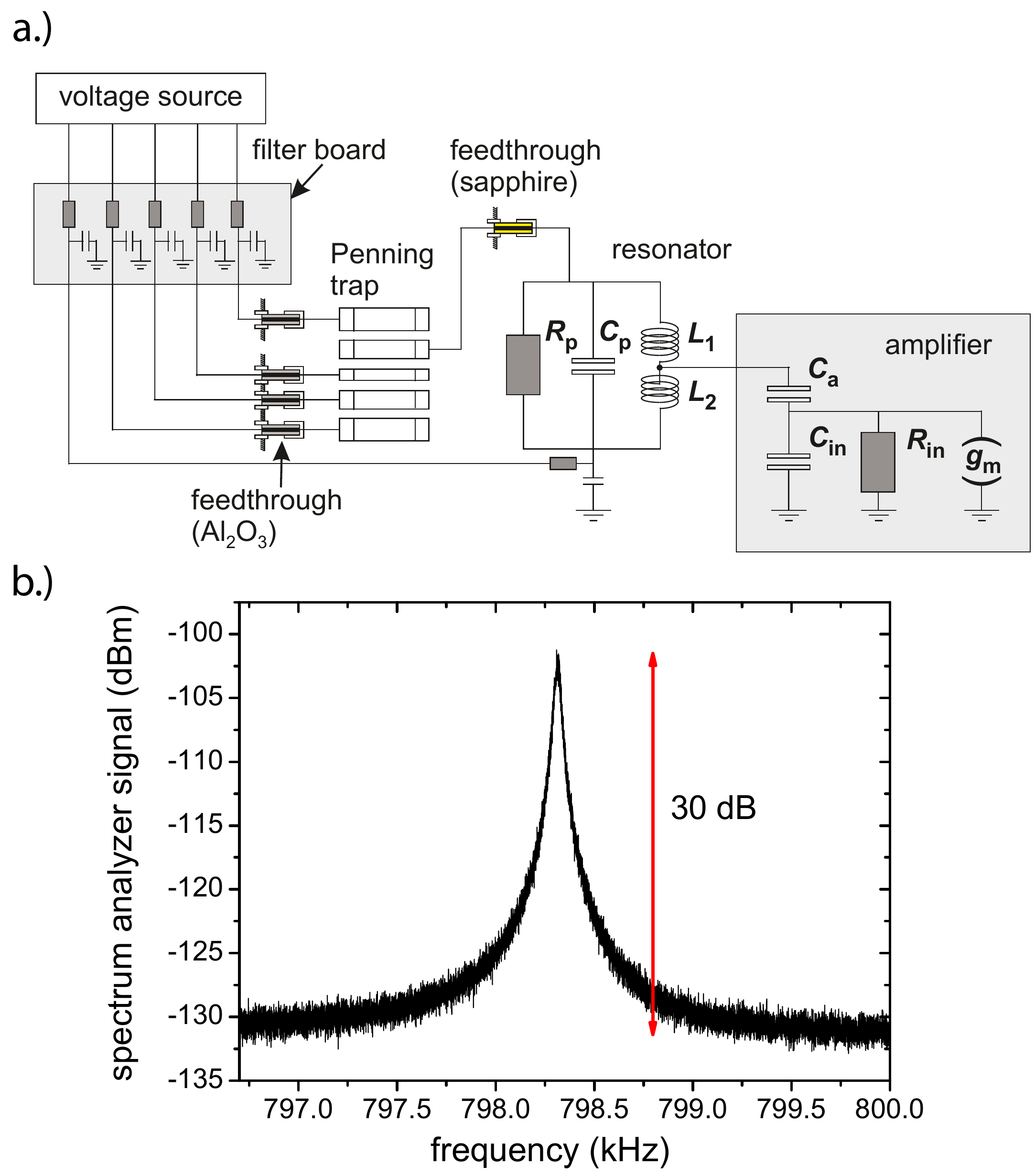}
		\caption{a.) Detailed layout including the input characteristics of the amplifier, voltage biasings to the trap electrodes from the voltage source, and possible lossy components which are considered to affect the effective parallel resistance of the whole system $R_{\textrm{eff}}$. b.) Recorded noise spectrum of the detector which is used for the reservoir trap (RT). The SNR is 10$\,$dB higher than our previous similar detection system. This allows much faster and more precise measurements of the axial frequency.}
	\end{center}
\end{figure}
Resonator and amplifier are coupled as described above and implemented into the Penning trap apparatus, as shown in FIG$.\,$4$\,$a.). The trap is located in a cryopumped vacuum chamber with a volume of 1.2$\,$l, the detectors are placed outside of this chamber. To connect the detectors to the trap electrode, cryogenic feedthroughs are used.  We found that alumina feedthroughs contribute parasitic losses at the frequencies of interest, which can be characterized by effective parallel resistances of $R_{\textrm{p,ft,A}}\approx120\,$M$\Omega$. To avoid these losses, we use feedthroughs based on sapphire dielectrics (Kyocera). For these parts we determined $R_{\textrm{p,ft,S}}>7\,$G$\Omega$.
To characterize the detectors we record time transients and perform fast Fourier transforms on the measured data. A recorded noise spectrum of the RT detector is shown in FIG$.\,$4$.\,$b.). From such spectra, the resonance frequency $\nu_{0}$ and the $Q$-value are extracted. The effective parallel resistance of the whole system $R_{\textrm{eff}}$ is calculated by applying the obtained $Q$-value, $\nu_{0}$, and the inductance $L$ to $R_{\text{eff}}=2\pi \nu_{0}LQ$. The obtained results are summarized in TABLE$\,$II. In case of the RT and the AT detectors, the measured values correspond exactly to the design specifications. Compared to our detection systems developed previously \cite{UlmerPRL}, the SNR is improved by up to 10$\,$dB. The major improvement is caused by using sapphire instead of alumina for the feedthroughs which increased input resistance, and slightly reduced equivalent input noise of the amplifier. This allows for faster and more precise determinations of the axial frequency, demonstrated by Eq.$\,$3.\\
The $Q$-value of the PT detector is approximately a factor of 3 smaller than expected. The details of the corresponding limitation have yet to be understood. The effective temperature $T_{z}$ of our detection systems is 7.8(1.2)$\,$K, determined as described in \cite{Geonium-lineshape}. This is close to the physical temperature of our apparatus.
\begin{table}[htb]
	\caption{Characterized parameters of the axial detection systems which are implemented in the BASE apparatus. Abbreviations: RT - Reservoir trap / PT - Precision trap / AT - Analysis trap / SNR - Signal-to-noise ratio / $\nu_{0}$ - Resonance frequency / $R_{\textrm{eff}}$ - Effective parallel resistance of the detector.}
  \begin{tabular}{|c||c|c|c|c|} \hline
     & SNR & $Q$-value & $\nu_{0}$ & $R_{\textrm{eff}}$ \\ \hhline{|=#=|=|=|=|}
    RT & 30$\,$dB & 20$\,$000 & 798$\,$kHz & 170$\,$M\textrm{$\Omega$} \\ \hline
    PT & 25$\,$dB & 6\,800 & 676$\,$kHz & 49.4$\,$M\textrm{$\Omega$} \\ \hline
    AT & 27$\,$dB & 26$\,$000 & 674$\,$kHz & 275$\,$M\textrm{$\Omega$} \\ \hline
  \end{tabular}
\end{table}

\section{VII. \textit{Q}-Tuner / Superconducting switch}
\begin{figure}[htbp]
	\begin{center}
		\includegraphics[width=8.5cm]{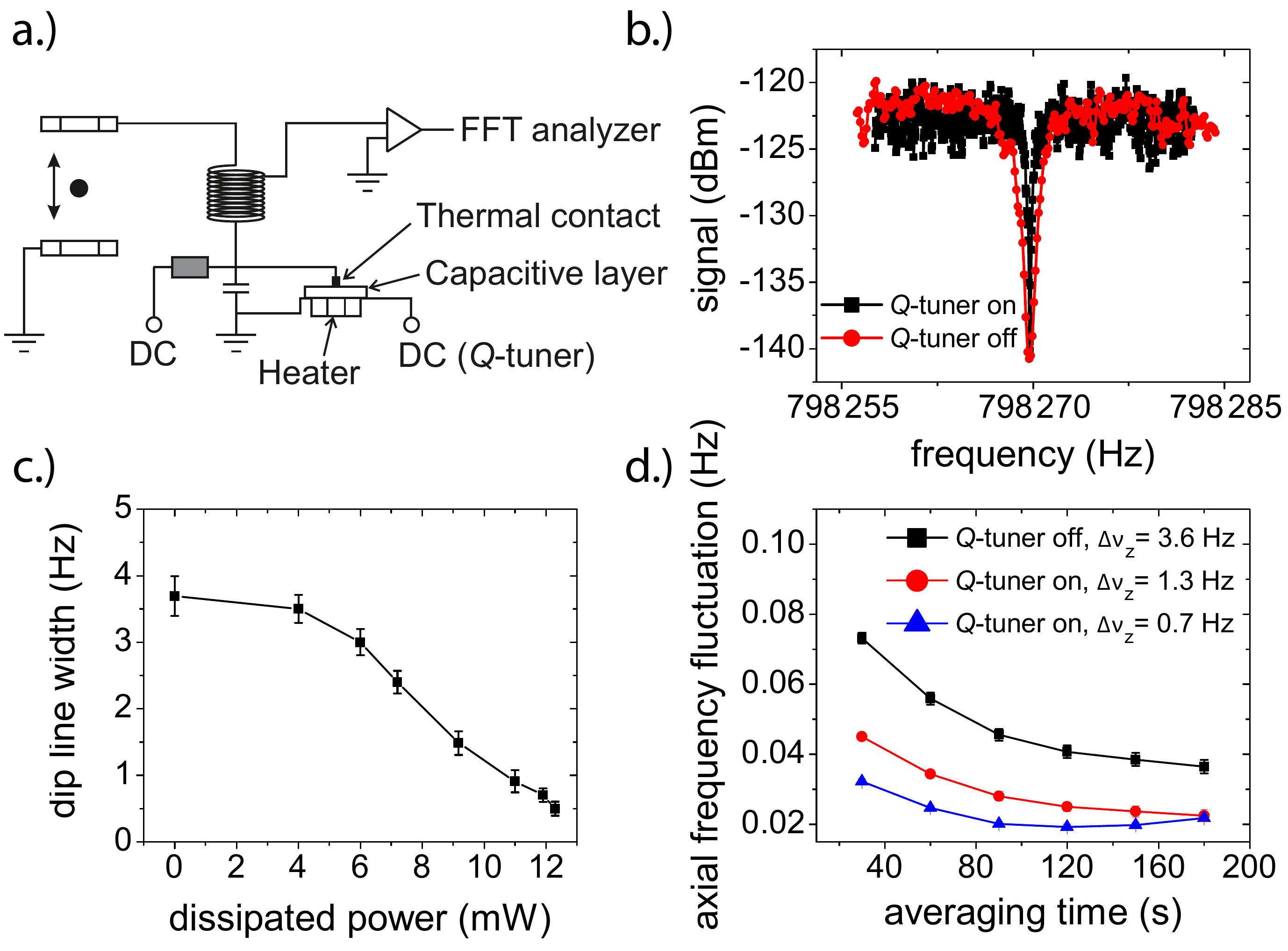}
		\caption{$Q$-tuner / superconducting switch. a.) working principle of the $Q$-tuner. Via a dielectric layer a heater is connected to a thermal link which is attached to the superconducting resonator. b.) Axial frequency spectra of a single trapped antiproton with $Q$-tuner off (red round data points) and $Q$-tuner on (black square data points). c.) Line width of the single antiproton dip as a function of the dissipated power. d.) Characterization of the axial frequency stability based on the Allan deviation evaluation, comparing three different conditions.}
	\end{center}
\end{figure}
The effective parallel resistance of the detector $R_{\textrm{eff}}$, together with the coupling strength $1/D$, defines the particle-detector interaction and therefore the line-width of the single-particle dip or the SNR in peak-detection methods. To achieve high frequency resolution in noise-dip based measurements, small $\Delta\nu_z$ and inherently weak particle-detector interaction is favorable, while peak detection requires strong particle-detector coupling \cite{Cornell}. To adjust the detector to the requirements of the respectively applied method, we developed a superconducting switch which allows continuous and sensitive tuning of $R_{\textrm{eff}}$.\\
The working principle of this switch is illustrated in FIG$.\,$5$\,$a.). A thermal conductor is attached via a galvanic connection to the cold end of the resonator. The other end of the conductor is attached to a metallized dielectric plate which has a small dielectric loss-angle $\delta$. A circuit which drives the heater is connected to the lower plate. By dissipating power at the heater, part of the superconducting coil is heated over the superconducting phase transition. This adds an additional series resistance to the resonator and $R_{\textrm{{eff}}}$ can be adjusted continuously, in a range of $\approx 0\,\Omega$ and 275$\,\text{M}\Omega$. This principle allows the continuous tuning of the particle-detector interaction. However, compared to GaAs-FET based $Q$-switches or active feedback techniques \cite{durso}, no parasitic parallel resistance or spurious thermal noise is added to the detection circuit.
FIG.$\,$5 b$.$) shows FFT spectra of a single trapped antiproton. The red round points show the $Q$-tuner switched off, the black square points represent a noise spectrum where a power of about 10$\,$mW was dissipated at the heater of the $Q$-tuner. This narrows the width of the axial dip of the antiproton to $\Delta\nu_z=1.0\,$Hz. In FIG$.\,$5 c.) $\Delta\nu_z$ is shown as a function of the power dissipated on the heater. Within a dissipation range of only 10$\,$mW, contributing only about 5$\,\%$ of the total power consumed by the 4$\,$K stage of our experiment, the line-width of the dip can be adjusted in a range between 3.5$\,$Hz and 0.7$\,$Hz. By increasing the power dissipated on the switch, the line-width of the dip could be reduced further, however without significant improvement of frequency resolution. FIG$.\,$5 d$.$) shows Allan deviations of the axial frequency stability, which is defined as the standard deviation of two subsequent axial frequency measurements $\sigma(\nu_{z,k}(t_a)-\nu_{z,k-1}(t_a))$, $t_a$ being the averaging time per frequency measurement. The black square markers represent the stability conditions achieved with the $Q$-tuner turned off. While recording the data-sets represented by the red round and the blue triangle data-points, the $Q$-tuner was adjusted to reduce the line-width of the single antiproton dip to $\Delta\nu_z=1.3\,$Hz (indicated by the red round plot) and $\Delta\nu_z=0.7\,$Hz (blue triangle plot). For averaging times $t_a<120\,$s, the reduction of the axial frequency fluctuation fits the theoretical expectation as described by Eq$.\,$(3). At $t_a>120\,$s, a random walk component, caused by drifts in the voltages which define the trapping potential, starts to contribute and modifies the $1/\sqrt{t_a}$ scaling. In total, however, for a wide range of averaging times the use of the $Q$-tuner reduces the axial frequency fluctuation $\propto\sqrt{\Delta\nu_{z}/t_a}$ as expected. Therefore, by using this device all types of Penning-trap experiments which rely on axial-dip measurements, such as proton and antiproton magnetic moment measurements \cite{ref10,Jack} and charge-to-mass ratio comparisons \cite{charge-to-mass ratio Nature}, as well as measurements of the $g$-factor of the electron bound to highly-charged ions, could potentially improve experimental precision when using such a device.

\section{VIII. Summary}
We have developed image current detection systems in toroidal design. The unloaded $Q$-values of these devices are up to $5 \times 10^{5}$. For our single antiproton spectroscopy experiments, we tune the detectors to $Q \approx 20\,000$ and signal-to-noise ratios of 30$\,$dB. In addition, a newly developed superconducting switch enables us to tune the resonance widths of single particle noise dips, which significantly reduces measured axial frequency fluctuations at low averaging times, which will improve experimental precision in our future planned proton-to-antiproton magnetic moment and charge-to-mass ratio comparisons.

\section{Acknowledgements}
We acknowledge support by the Antiproton Decelerator group, CERN's cryolab team, and all other CERN groups which provide support to Antiproton Decelerator experiments. We acknowledge financial support by RIKEN Initiative Research Unit Program, RIKEN President Funding, RIKEN Pioneering Project Funding, RIKEN FPR Funding, the RIKEN JRA Program, the Grant-in-Aid for Specially Promoted Research (grant number 24000008) of MEXT, the Max-Planck Society, the EU (ERC advanced grant number 290870-MEFUCO), the Helmholtz-Gemeinschaft, and the CERN Fellowship program.


\begin{thebibliography}{99}
\bibitem{ref1} K. Blaum, Phys. Rep. \textbf{425}, 1 (2006).
\bibitem{E.Myers} E. G. Myers, Int. J. Mass Spectrom. \textbf{349-350}, 107 (2013).
\bibitem{He-3} E. G. Myers, A. Wagner, H. Kracke, and B.A. Wesson, Phys. Rev. Lett. \textbf{114}, 013003 (2015).
\bibitem{ref2} L. Gastaldo, K. Blaum, A. Doerr, Ch. E. D\"ullmann, K. Eberhardt, S. Eliseev, C. Enss, Amand Faessler, A. Fleischmann, S. Kempf, M. Krivoruchenko, S. Lahiri, M. Maiti, Yu. N. Novikov, P. C.-O. Ranitzsch, F. Simkovic, Z. Szusc, and M. Wegner, J. Low Temp. Phys. {\bf 176}, 876 (2014).
\bibitem{ref3} S. Rainville, J. K. Thompson, E. G. Myers, J. M. Brown, M. S. Dewey, E. G. Kessler, Jr., R. D. Deslattes, H. G. B\"orner, M. Jentschel, P. Mutti, and D. E. Pritchard, Nature \textbf{438}, 1096 (2005).
\bibitem{ref4} A. Wagner, S. Sturm, F. K\"ohler, D. A. Glazov, A. V. Volotka, G. Plunien, W. Quint, G. Werth, V. M. Shabaev, and K. Blaum, Phys. Rev. Lett. \textbf{110}, 033003 (2013).
\bibitem{ref5} S. Sturm, F. K\"ohler, J. Zatorski, A. Wagner, Z. Harman, G. Werth, W. Quint, C. H. Keitel, and K. Blaum, Nature \textbf{506}, 467 (2014).
\bibitem{ref6} R. S. Van Dyck, Jr., P. B. Schwinberg, and H. G. Dehmelt, Phys. Rev. Lett. \textbf{59}, 26 (1987).
\bibitem{ref7} G. Gabrielse, A. Khabbaz, D. Hall, C. Heimann, H. Kalinowsky, and W. Jhe, Phys. Rev. Lett. \textbf{82}, 3198 (1999).
\bibitem{ref8}  J. DiSciacca and G. Gabrielse, Phys. Rev. Lett. \textbf{108}, 153001 (2012).
\bibitem{ref9} C. Smorra, K. Blaum, L. Bojtar, M. Borchert, K. A. Franke, T. Higuchi, N. Leefer, H. Nagahama, Y. Matsuda, A. Mooser, M. Niemann, C. Ospelkaus, W. Quint, G. Schneider, S. Sellner, T. Tanaka, S. V. Gorp, J. Walz, Y. Yamazaki, and S. Ulmer, Eur. Phys. J. ST \textbf{224}, 3055 (2015).
\bibitem{charge-to-mass ratio Nature} S. Ulmer, C. Smorra, A. Mooser, K. Franke, H. Nagahama, G. Schneider, T. Higuchi, S. V. Gorp, K. Blaum, Y. Matsuda, W. Quint, J. Walz, and Y. Yamazaki, Nature. \textbf{524}, 196 (2015).
\bibitem{ref10} A. Mooser, S. Ulmer, K. Blaum, K. Franke, H. Kracke, C. Leiteritz, W. Quint, C. C. Rodegheri, C. Smorra, and J. Walz, Nature \textbf{509}, 596 (2014).
\bibitem{image current} D. J. Wineland and H. G. Dehmelt, J. Appl. Phys. \textbf{46}, 919 (1975).
\bibitem{invariance} L. S. Brown and G. Gabrielse, Phys. Rev. A. \textbf{25}, 2423 (1982).
\bibitem{UlmerPRL}
S. Ulmer, C. C. Rodegheri, K. Blaum, H. Kracke, A. Mooser, W. Quint, and J. Walz, Phys. Rev. Lett. \textbf{106}, 253001 (2011).
\bibitem{MooserPRL}
A. Mooser, H. Kracke, K. Blaum, C. C. Rodegheri, W. Quint, S. Ulmer and J. Walz, Phys. Rev. Lett. \textbf{110}, 140405 (2013).
\bibitem{Bc2} Y. Shapira and L. J. Neuringer, Phys. Rev. \textbf{140}, A1638 (1965).
\bibitem{Stefan_RSI} S. Ulmer, H. Kracke, K. Blaum, S. Kreim, A. Mooser, W. Quint, C. C. Rodegheri, and J. Walz, Rev. Sci. Instrum. \textbf{80}, 123302 (2009).
\bibitem{Stefan cyclotron} S. Ulmer, K. Blaum, H. Kracke, A. Mooser, W. Quint, C. C. Rodegheri, and J. Walz, Nucl. Instr. Meth. Phys. Res. A \textbf{705}, 55 (2013).
\bibitem{PCB} Rogers corporation, RT/duroid 5870/5880 data sheet.
\bibitem{Geonium-lineshape} L. S. Brown, Ann. Phys. \textbf{159}, 62 (1985).
\bibitem{Cornell} E. A. Cornell, R. M. Weisskoff, K. R. Boyce, R. W. Flanagan, Jr., G. P. Lafyatis, and D. E. Pritchard, Phys. Rev. Lett. \textbf{63}, 1674 (1989).
\bibitem{durso} H. Dehmelt, W. Nagourney, and J. Sandberg, Proc. Nat. Acad. Sci. \textbf{83}, 5761 (1986).
\bibitem{Jack} J. DiSciacca, M. Marshall, K. Marable, G. Gabrielse, S. Ettenauer, E. Tardiff, R. Kalra, D. W. Fitzakerley, M. C. George, E. A. Hessels, C. H. Storry, M. Weel, D. Grzonka, W. Oelert, and T. Sefzick, Phys. Rev. Lett. \textbf{110}, 130801 (2013).
 \end{thebibliography}
\end{document}